\documentclass[english]{article}
\usepackage[utf8]{inputenc}
\usepackage[T1]{fontenc}
\usepackage{babel}
\usepackage{hyperref}
\usepackage{amsmath}
\usepackage{graphicx}
\usepackage{fancyhdr}
\newcommand{\scidatalogo}{\includegraphics[height=36pt]{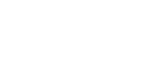}}
\newcommand{\overleaflogo}{\includegraphics[height=36pt]{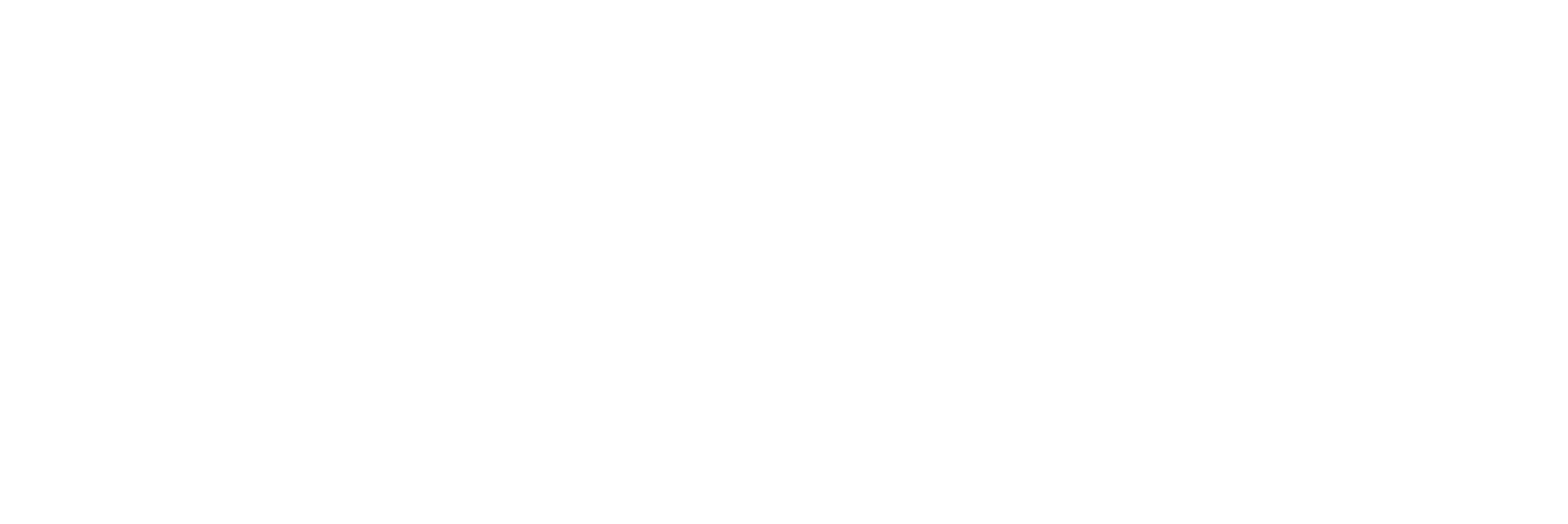}}
\begin{document}

\title{The Empusa code generator: bridging the gap between the intended and the actual content of RDF resources}

\author{Jesse C.J. van Dam\textsuperscript{1}, 
Jasper J. Koehorst\textsuperscript{1}, 
Peter J. Schaap\textsuperscript{1}, 
Maria Suarez-Diez\textsuperscript{1{*}}}

\maketitle
\thispagestyle{fancy}

1. Laboratory of Systems and Synthetic Biology, Wageningen University \& Research, Wageningen, 6708 WE, the Netherlands {*}corresponding author(s):
Maria Suarez-Diez (maria.suarezdiez@wur.nl)

\begin{abstract}
The RDF data model facilitates integration of diverse data available in structured and semi-structured formats. 
To obtain an RDF graph with a low amount of errors and internal redundancy, the chosen ontology must be consistently applied. However, with each addition of new diverse data the ontology must evolve thereby increasing its complexity, which could lead to accumulation of unintended erroneous composites. 
Thus, there is a need for a gatekeeping system that compares the intended content described in the ontology with the actual content of the resource.
 \newline
 Here we present Empusa, a tool that has been developed to facilitate the creation of composite RDF resources from disparate sources. Empusa can be used to convert a schema into an associated application programming interface (API) that can be used to perform data consistency checks and generates Markdown documentation to make persistent URLs resolvable. In this way, the use of Empusa ensures consistency within and between the ontology (OWL), the Shape Expressions (ShEx) describing the graph structure, and the content of the resource.
\end{abstract}

\section*{Background \& Summary}
Semantic Web technologies provide information retrieval and management systems to integrate heterogeneous data from disparate sources \cite{berners-lee}. The RDF data model is a W3C standard for storage of information in the form of self-descriptive Subject, Predicate and Object triples that can be linked in an RDF-graph \cite{brickley_rdf_2004, w3c_organisation_rdf_2014}. The use of retrievable controlled vocabularies enables integration of heterogeneous diverse data from different sources in a single repository and SPARQL can be used to query the so generated resources \cite{prudhommeaux_sparql_2008,w3c_organisation_sparql_2013}. 

By themselves, RDF graphs have no predefined structure nor a schema, and the structure of an RDF resource can vary as new triples are added. Therefore, a formal definition of the relations among the terms, called an ontology, is required to efficiently retrieve linked information from these resources. Structural information can be encoded using Web Ontology Language (OWL) files \cite{w3c_organisation_owl_2012}. RDFS is another, related, standard to define the structure of an RDF resource \cite{brickley_rdf_2014}. In this standard, each object can be defined as an instance of a class and each link as the realisation of a property. Shape Expressions (ShEx) is a standard to describe, validate and transform RDF data. One of the goals of this standard is to create an easy to read language for the validation of instance data \cite{solbrig_shape_2014,boneva_validating_2014,SHEX}. 

In previous work, we developed RDF2Graph, a tool to automatically recover the structure of an RDF resource and to generate a visualisation, ShEx file and/or an OWL ontology thereof \cite{RDF2Graph}. Application of RDF2Graph to resources providing data in the RDF data model in the life sciences domain such as Reactome, ChEBI, UniProt, or those transformed by the Bio2RDF project \cite{belleau_bio2rdf:_2008,croft_reactome_2014, hastings_chebi_2013, jupp_ebi_2014, TheUniProtConsortium2017} showed mismatches between the retrieved data structure and the one described in the OWL definition of the particular resource. The main reason for this lack of consistency is the flexibility provided by RDF: the data graph is a free format, the ontology defines the structure but does not enforce it. 

In the development of RDF resources, transformation of existing data into the RDF data model is often a source of errors such as typing errors in the predicates, instances with missing attributes, instances that did have a non-unique IRI, and instances that had no type defined, among others.  Development of tools that directly use the RDF data model as means to store their output may therefore be essential to unlock the potential of these technologies in the life sciences. An example of a such tool is the Semantic Annotation Platform with Provenance (SAPP) \cite{Koehorst2018}, that can automatically annotate genome sequences using standard tools and directly store the annotation results and their provenance in the RDF data model using the Genome Biology Ontology Language (GBOL) \cite{gbol}. Development of such tools would be greatly facilitated by supporting tools able to read an ontology definition and generate code that can be used for data generation, export and validation.

\begin{figure}
 \centering
 \includegraphics[width=1\textwidth]{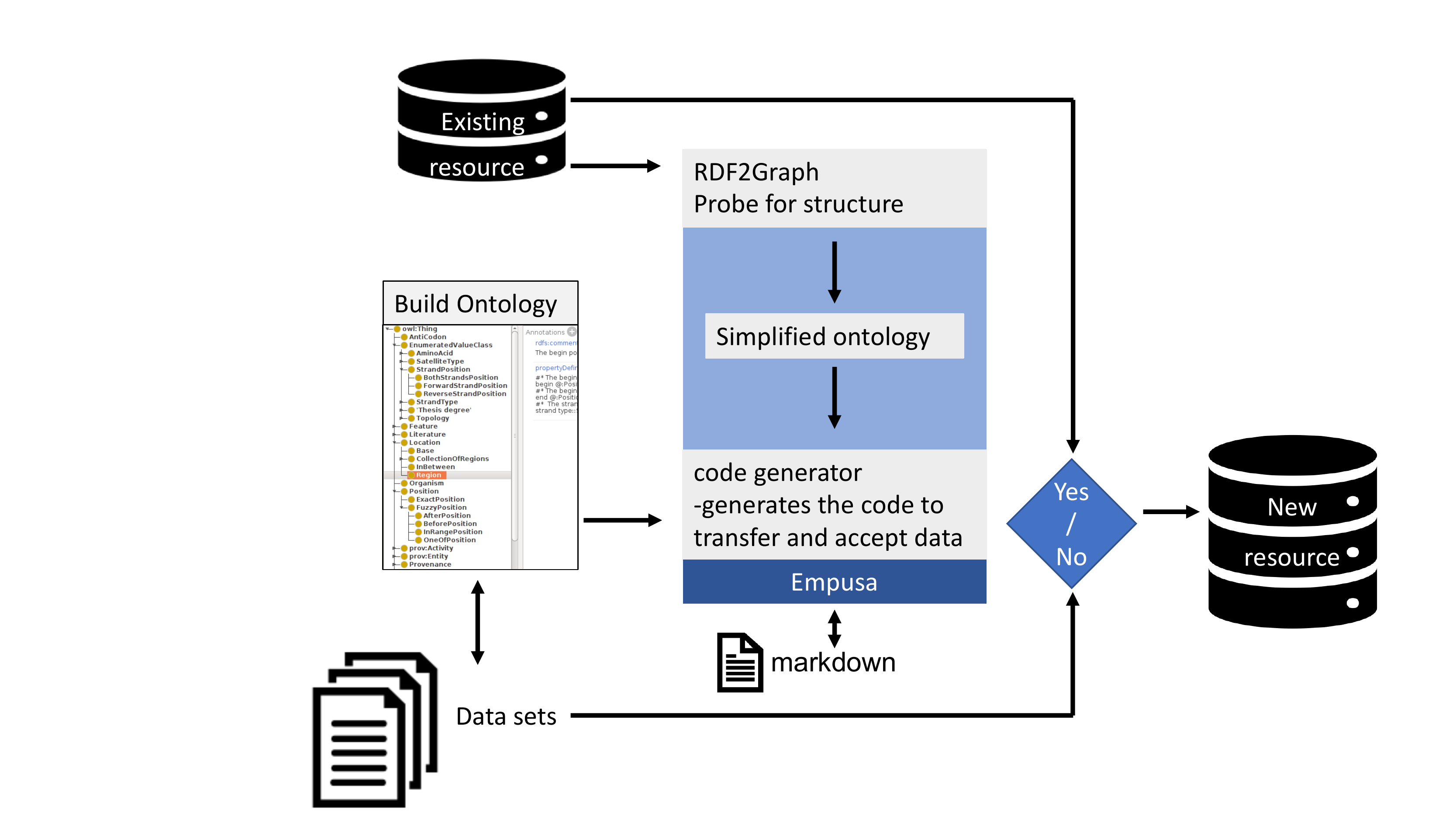}
 \caption{ \textbf{Simplified overview of the workflow to manage consistent integration of new diverse data with existing resources.}  Empusa enables error control as it compares the  intended content, described in the ontology,  with the actual content of the resource. For this, Empusa checks  whether or not \textit{Subjects} and \textit{Objects} have the properties that the ontology demands. Empusa builds upon RDF2Graph\cite{RDF2Graph}, a tool to automatically recover the structure of an RDF resource, to generate a visualisation, ShEx file, and/or an OWL ontology thereof.} 
 \label{fig:empusa_workflow}
\end{figure}

Here we present \textit{Empusa}, that has been developed to facilitate the creation of RDF resources, which are validated upon creation (figure \ref{fig:empusa_workflow}). Empusa uses an OWL and a simplified version of ShEx, defining an ontology, and generates an associated application programming interface (API) that can be used to perform data consistency checks. The use of Empusa ensures consistency within and between the ontology (OWL), the Shape Expressions (ShEx) describing the graph structure and the content of the resource. In addition, Markdown documentation is generated, making URLs related to the ontology resolvable \cite{gruber2004daring}.

\section*{Methods}
The input definition of Empusa is a combination between OWL and a simplified version of ShEx, which can be edited within Prot\'eg\'e \cite{PROTEGE}. The classes are defined in OWL, whereas the properties are defined in each class under the annotation property \textit{propertyDefinitions} encoded within a simplified format of the ShEx standard. Additionally predefined value sets can be defined by adding a subclass to the \textit{EnumeratedValueClass}. For instance a \textit{FileType} can only be one element of a predefined list (e.g. CSV,TXT,TSV). 

\begin{figure}
 \centering
 \includegraphics[width=1\textwidth]{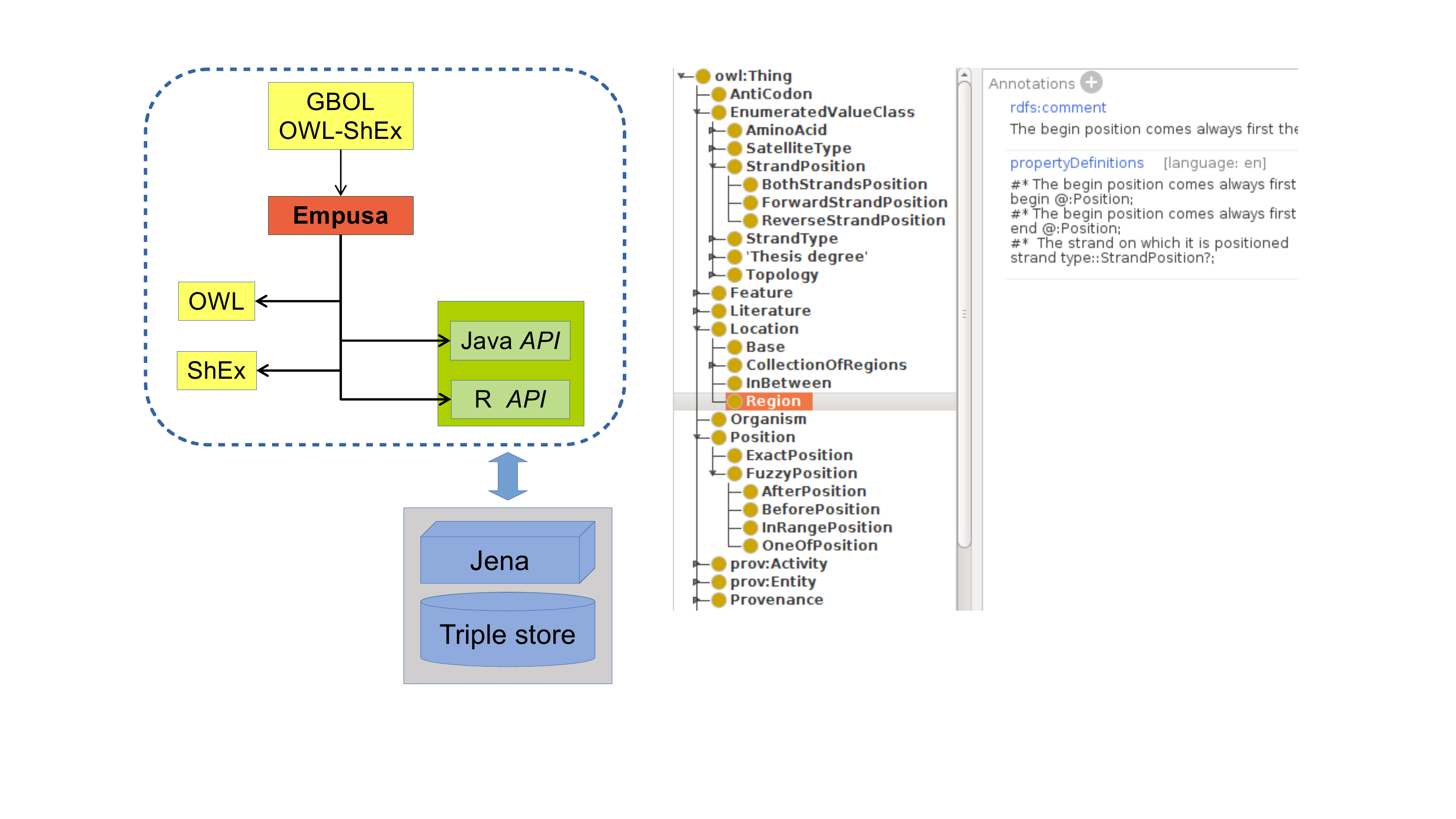}
 \caption{ \textbf{Empusa file definition}. \textit{Left}: The input definition file (combining 
 OWL and SHeX) is used to provide an ontology (here the GBOL\cite{gbol} ontology is used as example). Empusa generates as output: an OWL file definition, a ShEx file that can be used for instance validation, the corresponding documentation in Markdown format, and R and Java APIs.  \textit{Right} Example input file. Properties within a class can be defined with the \textit{propertyDefinitons} annotation property. As an example, the \textit{Region} class has been highlighted. Value sets (sub-ontologies) can be defined under the \textit{EnumeratedValueClass} class, for example the \textit{StrandPosition} value set. } 
 \label{fig:empusa_source}
\end{figure}

The RDFS standard is used to define the \textit{subClassOf} relationships between the classes, whereas the ShEx standard is used to define the properties of each class. Properties of the class are defined through the annotation property \textit{propertyDefinitions} as shown in figure \ref{fig:empusa_source}. For each property the multiplicity and the expected type of the target value can be defined. The multiplicity can either be: \textit{0..1} indicating that the property is optional and at most one reference is allowed; \textit{1..1} indicating that one and only reference is allowed; \textit{0..N} for optional properties with multiple allowed references; and \textit{1..N} for properties that must have at least one reference. The ‘=’ and ‘$\sim$’ sign can be used to define the references to be stored as an ordered or numbered list to ensure that the elements are numbered. Target value types can also be defined. The type of the target value can be either: A simple value (String, Integer or Double, among others); Another class (for example a Protein); Or an IRI, referencing an external resource or ontology or to a sub-ontology (value set). Within the ontology, sub- ontologies  (value sets) can be defined under the \textit{EnumeratedValue} class. Every sub-class of \textit{EnumeratedValue} class represents one sub ontology. All subsequent sub-classes are elements of the sub-ontology of which it is sub-classed from. A class/sub-class structure can be defined for these elements within the sub-ontology. 

The Empusa code generator uses this definition to generate: (i) An OWL file definition. It should be noted that the OWL file definition is generated as it remains general consensus within the field of semantics that these files are created for each ontology. (ii) A full ShEx file that can be used to validate a data set containing information that is encoded with the ontology. (iii) An R and Java API, which one can use to generate the data with the encoding of the defined ontology. This API ensures that the multiplicities and referenced types are correct and prevents many errors in the data export. 
(iv) A full documentation of the ontology based on \textit{mkdocs}. The rdfs:label and skos:description properties can be used within the ontology to add a description about the classes and a comment line above each property definition in the simplified ShEx definition and can be used to add a description to each property.


\subsection*{Code availability}
Empusa is written in Java with Gradle as build system. Empusa codebase is available at \url{http://www.gitlab.com/Empusa} under the MIT license. Documentation and tutorials can be found at associated website \url{http://empusa.org}. 

\section*{Discussion}
Empusa was developed primarily to help develop ontologies focusing on their function as a database schema for RDF resources. The design principles "modularity", "human readability", and "annotation" are followed to ensure that the so generated ontology can be easily extended \cite{LinkedData}. Empusa can automatically and consistently generate an OWL and a ShEx definition, ontology documentation in Markdown, an API, a JSON-LD framing file and a visualisation. Empusa uses parts of the RDF2Graph tool \cite{RDF2Graph} to generate a representation that can be subsequently used to generate a visualisation within Cytoscape \cite{Cytoscape}. This allows users to browse the complete ontology intuitively. 

Development of Empusa was closely related to the development of the GBOL stack \cite{gbol} and the associated tool SAPP \cite{Koehorst2018}. GBOL enables interoperable genome annotation, as it deploys and extends existing ontologies to represent genomic entities, their properties and associated provenance. The GBOL stack contains over 80.000 lines of R and Java code, OWL and ShEx definition files, and documentation files (mkdocs format). Generating such a large amount of code would entail 1 year of manual work (considering an efficiency of 50 lines per hour) \cite{Nawrocki}. 
Moreover, during the development of the GBOL ontology countless updates were made to correctly encapsulate all the data and associated provenance. Most of these updates were based on insights gained through the data encoding process. Manually updating the code, without using the supporting Empusa tool, would have entailed so much work that it would still be an on-going process. Thus, the Empusa code generator can serve to reduce the time (and costs) associated to development of ontologies and tools. 

In conclusion, the Empusa code generator can be used to develop new ontologies combined with automatic generation of API and documentation. This reduces the complexity and time to extend and develop ontologies and tools able to exploit the full potential of Semantic Web technologies for heterogeneous data integration. Moreover, Empusa enables the validation of the generated resources and the verification of the consistency of the exported data thereby bridging the gap between the intended and the actual content of RDF resources.

\section*{Acknowledgements}
This work has received funding from the Research Council of Norway, No. 248792 (DigiSal) and from the European Union FP7 and H2020 under grant agreements No. 305340 (INFECT), No. 635536 (EmPowerPutida), No. 634940 (MycoSynVac), No. 730976 (IBISBA 1.0), and the Netherlands Organisation for Scientific Research funded UNLOCK project (NRGWI.obrug.2018.005).
\newline

JvD was the primary developer of Empusa, explored the use cases and applications and drafted the manuscript.
JK  participated in code development and testing, explored the use cases and applications and revised the manuscript.
PS explored the use cases and applications and revised the manuscript.
MS-D explored the use cases and applications and revised the manuscript.
All authors critically read, revised and approved the manuscript. 
All authors had full access to the underlying code and data.




\section*{Competing financial interests}

The author(s) declare no competing financial interests.

\end{document}